\documentclass[conference]{IEEEtran}
\usepackage{amsmath,amssymb}
\usepackage{graphicx}
\usepackage{fancyvrb}
\usepackage{url}
\usepackage{microtype}
\usepackage{wrapfig}
\usepackage{hyperref}
\begin{document}
\bstctlcite{IEEEexample:BSTcontrol}% short references: first author et al.

\title{CAPER: Circuit-Aware Parameter-Efficient Reasoning in LLMs for Analog Verification}

\author{\IEEEauthorblockN{Saion K. Roy}
\IEEEauthorblockA{Northeastern University, Boston, MA, USA\\
sai.roy@northeastern.edu}}

\maketitle

\begin{abstract}
Large language models (LLMs) are increasingly used to generate analog circuits, yet every netlist must also be verified. We propose LLMs as verification assistants for analog circuits that flag unmet specifications and reliability issues that appear under process, voltage, and temperature (PVT) variation. These issues emerge from how circuit stages combine, a structure that generic parameter-efficient fine-tuning (PEFT) methods such as LoRA do not capture. To this end, we introduce CAPER, a circuit-aware PEFT method that adds a small branch alongside LoRA to read each stage's properties from the LLM and combine them from the relevant stages. In deciding whether a feedback loop is negative, LoRA and a high-rank adapter stay at chance, whereas CAPER answers every test case correctly. For comparator latches, where 11 of 90 LLM-written designs that work under nominal conditions fail under PVT variation, CAPER likewise learns to flag such latent failures, while generic adapters stay near chance. These results show that adding circuit knowledge to PEFT makes such analog verification checks learnable through fine-tuning.
\end{abstract}

\section{Introduction}
LLMs can now write analog netlists~\cite{lai2025analogcoder}, generate circuit topologies~\cite{gao2025analoggenie}, and answer circuit design questions. Every netlist, whether written by a designer or a model, must be verified against its specification. For analog circuits, this requires checking reliable operation across process, voltage, and temperature (PVT) variation, device mismatch, and the intended frequency and input ranges. Prior work has applied LLMs extensively to assertion-based verification for digital circuits~\cite{yan2025assertllm,kande2023llmassert}, whereas analog verification works have focused on checking netlist syntax and structure~\cite{chaudhuri2024spiced,ma2026netlistbench}. In this work, we propose using LLMs as analog verification assistants that review netlists against specifications, help designers identify functional errors, and flag reliability problems.

\begin{figure}[t]
\centering
\includegraphics{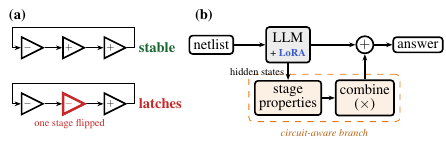}
\vspace{-0.25cm}
\caption{(a) Flipping one stage turns a stable loop into a latch. (b) CAPER adds a circuit-aware branch next to LoRA.}
\vspace{-0.25cm}
\label{fig:overview}
\end{figure}

These issues are hard to predict because they emerge from how circuit stages combine. Each stage has a local property, such as gain or polarity (whether it inverts the signal), and the circuit's behavior follows from how the stages connect. For example, at low frequencies, a feedback loop is negative when an odd number of its stages invert (Fig.~\ref{fig:overview}a). A single stage-polarity error can therefore turn negative feedback into positive feedback and cause the amplifier to latch. LLMs struggle with this reasoning: on feedback circuits in CircuitSense, none of eight leading models derives the transfer function correctly more than 40\% of the time~\cite{akbari2026circuitsense}. A verification assistant must therefore reason about how stages combine, not only about the individual components.

Adapting an LLM to a new task usually relies on parameter-efficient fine-tuning (PEFT), which keeps the pretrained model fixed and trains only a small number of added parameters. The most widely used PEFT method, LoRA~\cite{hu2022lora}, adds a small low-rank update to selected weight matrices, which lets the model adjust its behavior while training well under $1\%$ of its parameters. High-rank adapters such as QuanTA~\cite{chen2024quanta} learn a richer, higher-rank update. However, both are generic: neither encodes how the stages of a circuit combine. We therefore introduce CAPER (Circuit-Aware PEFT), which builds this combination directly into the adapter (Fig.~\ref{fig:overview}b). We validate CAPER on feedback polarity (Sec.~\ref{sec:fb}), and then test it on latent PVT failures in comparator latches (Sec.~\ref{sec:latent}).

\vspace{-0.1cm}
\section{CAPER: Circuit-Aware PEFT}
\label{sec:method}
Many verification questions reduce to combining stage-level facts. When the verdict depends on a product of many such facts, as for loop polarity, knowing only some of them gives no information about the answer. The training gradient therefore carries almost no learning signal about which stages to combine~\cite{shalev2017failures}. High-rank adapters such as QuanTA~\cite{chen2024quanta} add capacity, but capacity does not create this missing signal.

CAPER computes this combination explicitly in a small branch next to LoRA, in three steps: (1) linear probes read a local property of each stage or device from the LLM's hidden state, (2) a selection step picks the properties whose combination best explains the model's errors on a separate set of netlists, and (3) the branch merges the selected properties into a single feature. This feature enters the network through a small linear layer, initialized to zero, whose output is added to the final hidden state of the LoRA-adapted model just before the output layer (Fig.~\ref{fig:overview}b). Because this layer starts at zero, training begins exactly from plain LoRA, and the branch gains influence only when it lowers the loss.
 
\section{Verification Examples}
\label{sec:results}
\subsection{Feedback Polarity}
\label{sec:fb}
We generate SPICE netlists of 12-stage feedback loops in which each stage is a single NMOS transistor. A common-source (CS) stage takes the output at the drain and inverts the signal, whereas a source follower (SF) takes it at the source and does not. The model must decide whether the loop provides negative feedback, as in this abbreviated example:
\begin{Verbatim}[fontsize=\scriptsize, frame=single, framesep=2pt, samepage=true]
M1 n2 n1 vss vss nch W=2u L=0.18u      CS: inverts
M2 vdd n2 n3 vss nch W=2u L=0.18u      SF: follows
...   (9 more stages)
M12 n1 n12 vss vss nch W=2u L=0.18u    CS: closes loop
Is the feedback around the closed loop negative?
\end{Verbatim}
The full loop has six inverting stages, an even number, so the correct answer is ``no.'' Stage types are drawn at random, so chance accuracy is $50\%$, and we use 510 training, 510 validation, and 1,020 test loops.

We first probe a frozen CodeLlama-7B~\cite{roziere2023codellama}. Linear probes on an intermediate layer (layer 8 of 32) recover whether each of the 12 stages inverts with $100\%$ accuracy, yet they predict the polarity of the whole loop only at chance ($49.4\%$). The model thus knows every stage but not the rule that joins them. Because the loop gain is the product of the stage gains, loop polarity is a 12-input XOR of the stage inversions, which a linear readout cannot compute and gradient training struggles to learn~\cite{shalev2017failures}. Generic fine-tuning does not close this gap. With identical data, learning rate, target layers, and 600 steps, LoRA and the QuanTA-style adapter~\cite{chen2024quanta} stay at chance ($50.6\%$ and $50.4\%$), as does a tensor-train adapter~\cite{yang2024loretta}, even though the latter two have about $6\times$ more parameters than LoRA. LoRA stays at chance even when every stage polarity is written into the prompt, so neither added capacity nor explicit stage information lets a generic adapter find the rule.

CAPER succeeds because it adds this circuit knowledge to PEFT: \textit{the loop gain of a feedback circuit is the product of its stage gains}. Its selection step scores every candidate product against the model's errors and ranks the full 12-stage product first, so CAPER finds the loop without being given the topology. Thus, CAPER predicts every test loop correctly while adding only $\approx6\%$ to LoRA's trainable parameters. %\textit{A single piece of circuit knowledge, that gains multiply around a loop, thus turns a model that sees every stage into one that judges the whole loop correctly.}

%\vspace{-0.15cm}
\subsection{Latent Failures Under PVT Variation}
\label{sec:latent}
Our second check targets latent failures, that is, designs that pass at the nominal condition (typical corner, 1.8~V, 27~$^\circ$C) but fail under another PVT or mismatch condition. Of 90 comparator latches that a fine-tuned CodeLlama-7B wrote for specifications in the open-source SkyWater 130-nm process (SKY130) and that pass under nominal conditions, 11 fail elsewhere. One sizes its input pair at 12 and 6~$\mu$m instead of 10~$\mu$m and fails only at low supply and high temperature, where the imbalance acts as a built-in input offset. To study such failures at scale, we simulated 10,843 latches with input-pair asymmetries of up to $4\times$ under all PVT and mismatch conditions. The model sees the netlist and a question:
\begin{Verbatim}[fontsize=\scriptsize, frame=single, framesep=2pt, samepage=true]
XM1 outp outn vss vss nfet W=3.1u L=0.5u   cross-coupled
XM2 outp outn vdd vdd pfet W=7.7u L=0.5u   cross-coupled
...   (2 more cross-coupled, 2 dummies, 2 load caps)
XM5 outp vinp vss vss nfet W=16u  L=0.5u   input, vinp
XM6 outn vinn vss vss nfet W=4u   L=0.5u   input, vinn
Passes at nominal. Fails under other PVT or mismatch?
\end{Verbatim}
The vinp device is four times wider than the vinn device, so the answer is ``yes'': the latch fails at low V and high T.

Here, CAPER adds the rule: \textit{balancing the two halves of the latch}. A regenerative latch resolves toward its stronger side, so a width difference across a matched pair acts as an input offset. The selection step of CAPER picks the input pair first, correctly classifying $78.8\%$ of test latches, whereas LoRA and the QuanTA-style adapter stay near chance ($\approx53\%$). CAPER thus learns a reliability check that generic adapters miss.

\section{Conclusion \& Future Work}
This work thus introduces circuit-aware fine-tuning as a path toward LLM assistants that verify analog designs beyond nominal simulation. CAPER shows that building circuit knowledge into the adapter lets an LLM verify properties that generic PEFT fails to learn. In both checks, CAPER brings in a circuit-specific rule, the loop gain for feedback polarity and the balance of a matched pair for latent failures, and outperforms LoRA and a QuanTA-style adapter under the same fine-tuning. A natural next step is to scale CAPER to larger analog design libraries, such as AnalogGenie netlists~\cite{gao2025analoggenie}, and to a broader set of verification checks.

A second direction is quantum-inspired CAPER, aimed at the one step that does not scale. To find which stages to combine, CAPER scores every candidate combination, and the number of candidates grows as $2^n-1$ with the number of stages $n$, from 4,095 at 12 stages to over four billion at 32. These candidates share a structure that quantum systems represent naturally. Writing each stage polarity as $s_i=(-1)^{x_i}$ with $x_i\in\{0,1\}$, the loop polarity $\prod_{i\in S}s_i$ is the parity of the bits $x_i$, which a register with one qubit per stage reads out as the product of Pauli-$Z$ operators on the qubits in $S$. Parity is also efficiently learnable from noisy quantum examples~\cite{cross2015quantum}. Building on this construction, we plan to investigate quantum-inspired adapters that represent these high-order stage interactions compactly, without scoring every candidate, so CAPER can scale as a verification assistant for larger analog circuits.

\renewcommand{\footnotesize}{\scriptsize}% compact reference list
\bibliographystyle{IEEEtran}
{\raggedright\bibliography{refs_verified_no_links}}

\end{document}